\documentclass[10pt]{article}
\usepackage{amsmath}
\usepackage{amssymb}
\usepackage{amsfonts}
\newcommand{\be}{\begin{equation}}
\newcommand{\ee}{\end{equation}}
\newcommand{\bqn}{\begin{eqnarray}}
\newcommand{\eqn}{\end{eqnarray}}
\begin{document}

\title{The time evolution of an atom coupled to a thermal radiation field}
\author{G. Flores-Hidalgo\thanks{gflores@ift.unesp.br} \\
\textit{\small Instituto de Ci\^encias Exatas, Universidade
Federal de Itajub\'a, Av. BPS, Itajub\'a, MG, Brazil}}
\date{}
\maketitle

\begin{abstract}

We study the time evolution of an atom suddenly coupled to a
thermal radiation field. As a simplified model of the
atom-electromagnetic field system we use a system composed by a
harmonic oscillator linearly coupled to a scalar field in the
framework of the recently introduced dressed coordinates and
dressed states. We show that the time evolution of the thermal
expectation values for the occupation number operators depend
exclusively on the probabilities associated with the emission and
absorption of field quanta. In particular, the time evolution of
the number operator associated with the atom is given in terms of
the probability of remaining in the first excited state and the
decay probabilities from this state by emission of field quanta of
frequencies $\omega_k$. Also, it is showed that independent of the
initial state of the atom, it thermalizes with the thermal
radiation field in a time scale of the order of the inverse
coupling constant.

\vspace{0.34cm} \noindent PACS Number(s):~03.65Yz, 05.70.Ln,
05.30.Jp

\end{abstract}

\maketitle

\vskip2pc
\section{Introduction}

The study of systems out of thermal equilibrium has been since
long time ago one of the main active areas in physics. The actual
interest ranging from condensed mater physics to cosmology. In
most cases the interest is in the thermalization process, the
determination of the relevant time scales involved, together with
an understanding of the generation of entropy and particle
production in non equilibrium dissipative systems interacting with
an environment. However, despite the importance associate to these
processes, non equilibrium problems are still poorly understood
\cite{reviews}. The nontrivial non equilibrium dynamics of fields,
for instance, have diverse applications, finding use {\it e.g.} in
the studies concerning the recent experiments in
ultra-relativistic heavy-ion collision \cite{DCC}; applications to
the current problems of parametric resonance and particle
production in cosmology \cite{lindereh}; or in the context of the
recent studies involving the intrinsic dissipative nature of
interacting fields \cite{GR,BGR,BR}. In addition to that, typical
problems we have in mind to study are those related to the
nontrivial out-of-thermal equilibrium dynamics associated with
phase transitions in different physical systems. As a few examples
we may cite include the current applications to the study of
formation of Bose-Einstein condensates after a temperature quench
\cite{BEC}, or in the study of the dynamics of coupled fields
displaced from their ground states as determined by their free
energy densities \cite{RF}.  For recent attempts to solve some
related problems to the study of systems out of thermal
equilibrium see Refs.
\cite{salle,parisi,aarts1,aarts2,mosko,srednicki,gemmer,tasaki,scarani},
where use has been made of either analytical or numerical
approaches in the context of specific or general models. For
example, numerical studies have been performed in specific field
theoretical models in Refs. \cite{salle,parisi,aarts1,aarts2},
where the problems of equilibration and thermalization have been
studied. On the other hand in Ref. \cite{srednicki} the role of
chaos as a mechanism for quantum thermalization has been
considered. By supposing the validity of Berry's conjecture
\cite{berry} it has been showed that a rarified hard-spheres  gas
approaches a Maxwell-Boltzmann, Bose-Einstein or Fermi-Dirac
distribution according on wether the wave functions are taken to
be non-symmetric, completely symmetric or completely antisymmetric
functions of the particle position.

In recent works, in analogy with the renormalized fields in
quantum field theory, the concepts of dressed coordinates and
dressed states have been introduced
\cite{adolfo1,adolfo2,gabriel}. These concepts have been
introduced in the context of an atom, approximated by an harmonic
oscillator, linearly coupled to a scalar field, the whole system
being confined in a spherical cavity of diameter $L$. In terms of
dressed coordinates, dressed states have been defined as the
physically measurable states. The dressed states having the
physical correct property of stability of the oscillator (atom)
ground state in the absence of field quanta (the quantum vacuum).
For a recent clear explanation see Ref. \cite{yony}. Also, the
formalism showed to have the technical advantage of allowing an
exact computation of the probabilities associated with the
different oscillator (atom) radiation processes \cite{casana}. For
example, we obtained easily the probability of the atom to decay
spontaneously from the first excited state to the ground state for
arbitrary coupling constant, weak or strong and for arbitrary
cavity size. For weak coupling constant and in the continuum limit
$L\to\infty$ we obtained the old know result: $e^{-\Gamma t}$
\cite{adolfo1}. Also, considering a cavity of sufficiently small
radius \cite{adolfo2}, the method accounted for, the
experimentally observed, inhibition of the spontaneous decaying
processes of the atom \cite{hulet,haroche2}. In Refs.
\cite{nonlinear,yony} the concept of dressed coordinates and
states have been extended to the case in which nonlinear
interactions between the oscillator and the field modes are taken
into account. Furthermore, in Ref. \cite{eletromag} we considered
the oscillator electromagnetic field interaction model and in Ref.
\cite{casana2} dressed coordinates and states have been introduced
in the path integral formalism.

The aim of the present work is to study the thermalization process
in the framework of the afore mentioned dressed coordinates and
states. The physical situation that we have in mind is an atom
(approximated by an harmonic oscillator) initially in an arbitrary
state, suddenly coupled to a thermal radiation field (approximated
by an infinite set of harmonic oscillators at thermal
equilibrium).  Then, the purpose is to study the time evolution of
this initial state. Fundamental questions that we have to solve
are: the atom reaches a final equilibrium state? and if this is
the case, what is the meaning of this final equilibrium state? In
relation with these questions, it is also important to know the
time necessary for the atom to thermalize with the thermal
radiation field. By solving exactly this model we expect to gain
some insight to solve more complicated problems.

As already stated, in this paper we will treat a specific model
for an atom coupled to a thermal radiation field. Initially the
system is described by a density operator of the form

\be
\hat{\rho}=\hat{\rho}_0\otimes\hat{\rho}_\beta\;,
\label{aa1}
\ee
where $\hat{\rho}_0$ is the density operator for the atom, that
can be in an arbitrary pure or mixed state and $\hat{\rho}_\beta$
is the density operator for the radiation field at thermal
equilibrium at some given temperature $\beta^{-1}$. We specify
below the form of $\hat{\rho}_\beta$. At some time, that we take
as $t=0$, the atom is suddenly coupled to the thermal radiation
field, afterwards (the density operator of ) the total system
evolves according to the Liouville-Von Neumann equation.  An
equivalent description is to maintain constant the density
operator and take the operators (related to the physical
observables) as time dependent. Then, these operators evolve in
time according to the Heisenberg equation of motion

\be
\frac{\partial}{\partial t}\hat{O}(t)=i[\hat{H}, \hat{O}(t)]\;,
\label{a2}
\ee
where $\hat{O}(t)$ is a time dependent operator associated to some
physical observable and $\hat{H}$ is the Hamiltonian for the
atom-electromagnetic field system. As a model for this system we consider
the one with Hamiltonian given by

\begin{equation}
\hat{H}=\frac{1}{2}\left(\hat{p}_{0}^{2}+\omega_{0}^{2}\hat{q}_{0}^{2}\right)
+\frac{1}{2}\sum_{k=1}^{N}\left(\hat{p}_{k}^{2}+\omega_{k}^{2}\hat{q}_{k}^{2}
-2c_{k}\hat{q}_{k}\hat{q}_0\right)+
\frac{1}{2}\sum_{k=1}^N\frac{c_k^2}{\omega_k^2}\hat{q}_0^2\;,
\label{a3}
\end{equation}
where the limit $N\to\infty$ is understood, the subscript $0$
refers to the atom approximated by an harmonic oscillator of frequency
$\omega_0$, and
$k=1,2,...N$ refer to the harmonic field modes. Also we take
$\omega _{k}=2\pi /L$, $c_{k}=\eta \omega
_{k}$, $\eta = \sqrt{2g\Delta \omega }$ and $\Delta \omega =\omega
_{k+1}-\omega _{k}= 2\pi /L$, where $g$ is a frequency dimensional
coupling constant. At the end we will take the continuum limit
$L\to\infty$. The last term in Eq. (\ref{a3}) assures
the  positiveness of the Hamiltonian and it can be
seen as a frequency renormalization of the harmonic oscillator
\cite{thirring,weiss}.

A similar model to the one given by Eq. (\ref{a3}) has been used repeatedly from
time to time as a simplified model to describe the quantum Brownian
motion \cite{feynman,ulersma,caldeira,zurek1}, the decoherence problem
and other related problems \cite{zurek2,paz}. However, in all these previous
works no use has been made of the dressed coordinates. As we explain in
next section, when considering the Hamiltonian given by Eq. (\ref{a3}) as
the one for an atom-field electromagnetic field system, the introduction of
dressed (renormalized) coordinates will be necessary in order to guarantee
the stability of the atom ground state in the absence of field quanta.

Along this paper we use natural units $\hbar=c=k_B=1$.

\section{Dressed (renormalized) coordinates and the dressed density operator}
To make this paper self contained in this section we define  what
has been called  dressed coordinates and dressed states in Refs.
\cite{adolfo1,adolfo2,gabriel}. To understand the necessity of
introducing dressed coordinates in the system atom-electromagnetic
field system described by Hamiltonian (\ref{a3}), take $c_k=0$. In
this case the resulting free Hamiltonian admits the following
eigenfunctions,

\bqn
\psi_{n_0n_1...n_N}(q)&\equiv&
\langle q|n_0,n_1,...,n_N\rangle
\nonumber\\
&=&\prod_{\mu=0}^N\left[\left(\frac{\omega_\mu}{\pi}\right)^{1/4}
\sqrt{\frac{2^{-n_\mu}}{n_\mu!}}H_{n_{\mu}}(\sqrt{\omega_{\mu}}q_\mu)
e^{-\frac{1}{2}\omega_\mu q_\mu^2}\right]\;.
\label{e8a}
\eqn
The physical meaning of $\psi_{n_0n_1...n_N}(q)$ in this case is
clear, it represents the atom in its $n_0$-th excited level and
$n_k$ photons of frequencies $\omega_k$. Now, consider the state
$\psi_{n_00...0}(q)$: the excited atom in the quantum vacuum. We
know from experience that any excited level of the atom is
unstable. The explanation of this fact is that the atom is not
isolated from interacting with the quantum electromagnetic field.
This interaction in our model is given by the linear coupling of
$q_0$ with $q_k$. Obviously, when we take into account this
interaction any state of the type $\psi_{n_00...0}(q)$ is rendered
unstable. But there is a problem, the state $\psi_{00...0}(q)$,
that represents the atom in its ground state and no photons, is
also unstable contradicting the experimental fact of the stability
of the atom ground state. What is wrong? The first thing that
cames in our mind is to think that the model given by Eq.
(\ref{a3}) is wrong. Certainly, we know that the correct theory to
describe this physical system is quantum electrodynamics. On the
other hand such a description could be extremely complicated. If
we aim to  maintain the model as simple as possible and still
insist in describing it by the Hamiltonian given in Eq. (\ref{a3})
what we can do in order to take into account the stability of the
atom ground state? The answer lies in the spirit of the
renormalization program in quantum field theory: the coordinates
$q_\mu$ that appear in the Hamiltonian  are not the physical ones,
they are bare coordinates. We introduce dressed (or renormalized)
coordinates, $q_0'$ and $q_k'$, respectively for the dressed atom
and the dressed photons. We define these coordinates as the
physically meaningful ones. In terms of these coordinates we
define the dressed states by

\bqn
\psi_{n_0n_1...n_N}(q')&\equiv&
\langle q'|n_0,n_1,...,n_N\rangle_d \nonumber\\
&=&\prod_{\mu=0}^N\left[\left(\frac{\omega_\mu}{\pi}\right)^{1/4}
\sqrt{\frac{2^{-n_\mu}}{n_\mu!}}H_{n_{\mu}}(\sqrt{\omega_{\mu}}q_\mu')
e^{-\frac{1}{2}\omega_\mu (q_\mu')^2}\right]
\label{e8b}
\eqn
where the subscript $d$ means dressed state. The dressed states
given by Eq. (\ref{e8b}) are defined as the physically measurable
states and describe in general, the physical atom in the $n_0$-th
excited level and $n_k$ physical photons of frequencies
$\omega_k$. Obviously, in the limit in which the coupling constant
$c_k$ vanishes the renormalized coordinates $q_\mu'$ must approach
the bare coordinates $q_\mu$. Now, in order to relate the bare and
dressed coordinates we have to use the physical requirement of
stability of the dressed ground state. The dressed ground state
will be stable if it is defined as eigenfunction of the
interacting Hamiltonian given by Eq. (\ref{a3}). Also the dressed
ground state must be the one of minimum energy, that is, it must
be defined as being identical (or proportional) to the ground
state eigenfunction of the interacting Hamiltonian. From this
definition, one can construct the dressed coordinates in terms of
the bare ones.  Then, the first step in order to obtain the
dressed coordinates is to solve for the ground state eigenfunction
of the Hamiltonian given in Eq. (\ref{a3}). This bilinear
Hamiltonian can be diagonalized by introducing normal coordinates
and momenta $\hat{Q}_r$ and $\hat{P}_r$,
\begin{equation}
\hat{q}_{\mu }=\sum_{r=0}^{N}t_{\mu }^{r}\hat{Q}_{r}\;,
~~~\hat{p}_{\mu }=\sum_{r=0}^{N}t_{\mu}^{r}\hat{P}_{r}\;,~~~\mu =(0,k)\;,
\label{transf}
\end{equation}
where $\{t_{\mu }^{r}\}$ is an orthonormal matrix whose elements are given by
\cite{gabrielrudnei},
\begin{equation}
t_{k}^{r}=\frac{c_{k}}{(\omega_{k}^{2}-\Omega_{r}^{2})}t_{0}^{r}\;,
~~~~~~~~~~~ t_{0}^{r}= \left[1+\sum_{k=1}^{N}\frac{c_{k}^{2}}
{(\omega_{k}^{2}-\Omega_{r}^{2})^{2}}\right]^{-\frac{1}{2}}
\label{tkrg1}
\end{equation}
with $\Omega_r$ being the normal frequencies corresponding to the
collective modes of the coupled system and given as solutions of the
equation
\begin{equation}
\omega_0^2-\Omega_r^2=\sum_{k=1}^N\frac{c_k^2\Omega_r^2}
{\omega_k^2(\omega_k^2-\Omega_r^2)}\;.
\label{la5b}
\end{equation}
In terms of normal coordinates and momenta the Hamiltonian given by Eq. (\ref{a3})
reads as
\begin{equation}
\hat{H}=\frac{1}{2}\sum_{r=0}^{N}(\hat{P}_{r}^{2}+\Omega_{r}^{2}\hat{Q}_{r}^{2})\;,
\label{diagonal}
\end{equation}
then, the eigenfunctions of the Hamiltonian are given by
\begin{eqnarray}
\phi_{n_{0}n_{1}...n_N}(Q)&\equiv &
\langle Q|n_{0},n_{1},...,n_N\rangle_c
\nonumber \\
&=&\prod_{r=0}^N\left[\left(\frac{\Omega_r}{\pi}\right)^{1/4}
\sqrt{\frac{2^{-n_r}}{n_r!}}H_{n_{r}}(\sqrt{\Omega_{r}} Q_{r})
e^{-\frac{1}{2}\Omega_r^2}\right]\;,
\label{autofuncoes}
\end{eqnarray}
where the subscript $c$ means collective state. Now, using the
definition of the dressed coordinates:
$\psi_{00...0}(q')\propto\phi_{00...0}(Q)$ and using Eqs.
(\ref{e8b}) and (\ref{autofuncoes}) we get
$e^{-\frac{1}{2}\sum_{\mu=0}^N\omega_\mu (q_\mu')^2}=
e^{-\frac{1}{2}\sum_{r=0}^N\Omega_r Q_r^2}$, from which the
dressed coordinates are obtained as
\begin{equation}
q_\mu'=\sum_{r=0}^N\sqrt{\frac{\Omega_r}{\omega_\mu}}t_\mu^rQ_r\;.
\label{dress}
\end{equation}
%

\subsection{The dressed density operator}
If no use is made of the dressed coordinates and states, the
density operator for the radiation field at thermal equilibrium in
Eq. (\ref{aa1}) would be given by

\be
\hat{\rho}_\beta=
Z_\beta^{-1}\exp\left[-\beta\sum_{k=1}^N\omega_k\left(\hat{a}_k^\dag\hat{a}_k+
\frac{1}{2}\right)\right]\;,
\label{q2}
\ee
where $\hat{a}_k$ and $\hat{a}_k^\dag$ are annihilation and creation operators
and given by

\bqn
\hat{a}_\mu&=&
\frac{1}{\sqrt{2\omega_\mu}}\hat{p}_\mu-i\sqrt{\frac{\omega_\mu}{2}}\hat{q}_\mu
\label{a1a}\\
\hat{a}_\mu^\dag&=&\frac{1}{\sqrt{2\omega_\mu}}\hat{p}_\mu
+i\sqrt{\frac{\omega_\mu}{2}}\hat{q}_\mu\;.
\label{q4a}
\eqn
In Eq. (\ref{q2})  $Z_\beta=\prod_{k=1}^Nz_\beta^k$, is the partition function
of the thermal radiation field, where

\be
z_\beta^k=
{\rm Tr}_k\left[e^{-\beta\omega_k\left(\hat{a}_k^\dag\hat{a}_k+1/2\right)}\right]
=\frac{1}{2\sinh\left(\frac{\beta_k\omega_k}{2}\right)}\;.
\label{pfun}
\ee
Also, the density operator $\hat{\rho}_0$ for the atom  would be
written in terms of the coordinates $q_0$.

However, as explained above, in the context of an
atom-electromagnetic field system and described by Hamiltonian
given by Eq. (\ref{a3}) it is necessary to redefine what the
physical coordinates are for the atom and field modes. Then,
instead of the density operator given by Eq. (\ref{q2}), we have
to consider the one written in terms of dressed coordinates
$q_k'$, as the physically density operator for the radiation field
at thermal equilibrium,

\be
\hat{\rho}_\beta=Z_\beta ^{-1}
\exp\left[-\beta\sum_{k=1}^N\omega_k\left(\hat{a}_k'^{\dag}\hat{a}_k'+\frac{1}{2}.
\right)\right]
\label{ec6}
\ee
where $\hat{a}_k'$ and $\hat{a}_k'{^\dag}$ are dressed annihilation and creation
operators and given in terms of the dressed coordinates $q_k'$ by

\bqn
\hat{a}_\mu'&=&
\frac{1}{\sqrt{2\omega_\mu}}\hat{p}_\mu'-i\sqrt{\frac{\omega_\mu}{2}}\hat{q}_\mu'
\label{a1}\\
\hat{a}_\mu'^\dag&=&\frac{1}{\sqrt{2\omega_\mu}}\hat{p}_\mu'
+i\sqrt{\frac{\omega_\mu}{2}}\hat{q}_\mu'\;,
\label{q4}
\eqn
where in position representation
$\hat{p}_\mu'=-i\frac{\partial}{\partial q_\mu'}$.
Also, the density operator for the atom must be taken as the one
written in terms of the dressed coordinate $q_0'$.

Now, we are ready to study the time evolution of thermal
expectation values for relevant physical operators. We will be
mainly interested in the present work in the study of the time
evolution of the thermal expectation value of the time dependent
number occupation operator associated with the dressed oscillator
(the atom) $\hat{a}_0'{^\dag(t)}\hat{a}_0'(t)$.

\section{The thermalization process}

We state the thermalization problem as follows:{\it i)} the
initial state given by Eq. (\ref{a1}) will evolve in time to a
final equilibrium state? and {\it ii)} if the system evolves to a
final equilibrium state, is this an state of thermal equilibrium?. Also
we would like to know the mean time necessary for the system to reach
a final thermal equilibrium state.

Since any operator can be written in terms of annihilation and
creation operators, it will be sufficient to solve for the time
dependent annihilation and creation operators in order to solve
the out of thermal equilibrium problem. Using the Heisenberg
equation of motion, Eq. (\ref{a2}), we have for the time dependent
annihilation operator $\hat{a}_\mu'(t)$,

\be
\frac{\partial}{\partial t}\hat{a}_\mu'(t)=i[\hat{H},\hat{a}_\mu'(t)]
\label{h1}
\ee
and a similar equation for $\hat{a}_\mu'{^\dag}(t)$.
Obviously at $t=0$, $\hat{a}'_\mu(0)$, is given by Eq. (\ref{a1}). This
equation can be written, using Eqs. (\ref{transf}) and (\ref{dress}), as

\be
\hat{a}'(0)=\sum_{r,\nu=0}^N
\left(\frac{t_\mu^rt_\nu^r}{\sqrt{2\Omega_r}}\hat{p}_\nu-i
\sqrt{\frac{\Omega_r}{2}}t_\mu^rt_\nu^r\hat{q}_\nu\right)\;.
\label{h2}
\ee
In order to solve Eq. (\ref{h1}) we write $\hat{a}_\mu'(t)$ as
\be
\hat{a}_\mu'(t)=\sum_{\nu=0}^N\left(B(t)_{\mu\nu}\hat{p}_\nu+
\dot{B}_{\mu\nu}(t)\hat{q}_\nu\right)\;,
\label{h3}
\ee
where $B(t)_{\mu\nu}$ is a time dependent {\it c}-number and the
dot means derivative with respect to time. Replacing Eqs. (\ref{a3}) and
Eq. (\ref{h3}) in Eq. (\ref{h1}), working the
commutators and identifying identical operators in both sides of
the resultant equation, we obtain the following coupled equations
for $B_{\mu\nu}(t)$

\be
\ddot{B}_{\mu 0}(t)+
\left(\omega_0^2+\sum_{k=1}^N\frac{c_k^2}{\omega_k^2}\right)B_{\mu 0}(t)
-\sum_{k=1}^Nc_kB_{\mu k}(t)=0
\label{q7}
\ee
and

\be
\ddot{B}_{\mu k}(t)+\omega_k^2B_{\mu k}(t)-c_kB_{\mu 0}(t)=0\;.
\label{q8}
\ee
Note that above equations are identical to the classical equations of motion for
the bare coordinates $q_\mu$ that can be obtained using the Hamilton
equations of motion for the Hamiltonian given by Eq. (\ref{a3}).
Then we can decouple Eqs. (\ref{q7}) and (\ref{q8})
with the same matrix $\{t_\mu^r\}$ that diagonalizes the Hamiltonian (\ref{a3}),
that is, we can write for  $B_{\mu\nu}(t)$,

\be
B_{\mu\nu}(t)=\sum_{r=0}^Nt_\mu^rC_\nu^r(t)
\label{ec7}
\ee
and replacing the above equation in Eqs. (\ref{q7}) and (\ref{q8}), these
equations decouple into

\be
\ddot{C}_\mu^r(t)+\Omega_r^2C_\mu^r(t)=0\;,
\label{ex1}
\ee
from which we obtain $C_\mu^r(t)=a_\mu^re^{i\Omega_r t}+
b_\mu^re^{-i\Omega_r t}$. Then, substituting this expression in Eq.
(\ref{ec7}) we obtain

\be
B_{\mu\nu}(t)=\sum_{r=0}^Nt_\nu^r\left(a_{\mu}^re^{i\Omega_r t}+
b_{\mu}^re^{-i\Omega_r t}\right)\;.
\label{q9}
\ee
The time independent coefficients $a_{\mu}^r$ and $b_{\mu}^r$ are
determined by the initial conditions at $t=0$ for $B_{\mu\nu}(t)$
and $\dot{B}_{\mu\nu}(t)$. From Eqs. (\ref{h2}) and (\ref{h3}) we
find that these initial conditions are

\bqn
B_{\mu\nu}(0)&=&\sum_{r=0}^N\frac{t_\mu^rt_\nu^r}{\sqrt{2\Omega_r}}\;,
\label{q11}\\
\dot{B}_{\mu\nu}(0)
&=&-i\sum_{r=0}^N\sqrt{\frac{\Omega_r}{2}}t_\mu^rt_\nu^r\;.
\label{q12}
\eqn
Using the above initial conditions in Eq. (\ref{q9}) and the
orthonormality property of the matrix $\{t_\mu^r\}$ we obtain
$a_\mu^r=0$ and $b_\mu^r=\frac{t_\mu^r}{\sqrt{2\Omega_r}}$.
Replacing these values in Eq. (\ref{q9}) we get

\be
B_{\mu\nu}(t)=\sum_{r=0}^N\frac{t_\mu^rt_\nu^r}{\sqrt{2\Omega_r}}e^{-i\Omega_r t}\;.
\label{q13}
\ee
Using Eq. (\ref{q13}) in Eq. (\ref{h3}) we can get easily,

\be
\hat{a}_\mu'(t)=
\sum_{\nu=0}^Nf_{\mu\nu}(t)\hat{a}_\nu'\;, \label{q14}
\ee
where

\be
f_{\mu\nu}(t)=\sum_{r=0}^N t_\mu^rt_\nu^re^{-i\Omega_r t}\;.
\label{ex2}
\ee

Now, we can compute the time evolution of the expectation value
corresponding to the dressed occupation number operator
$\hat{n}_\mu(t)=\hat{a}_\mu'^\dag(t)\hat{a}_\mu'(t)$,

\be
n_\mu(t)={\rm Tr}\left[\hat{a}_\mu'^\dag(t)\hat{a}_\mu'(t)
\hat{\rho}_0\otimes\hat{\rho}_\beta\right]\;. \label{q15} \ee
where $\hat{\rho}_0$ is the dressed density operator corresponding to the atom
and $\hat{\rho}_\beta$ is the dressed density operator for the thermal
radiation field and given
by Eq. (\ref{ec6}). To compute the trace in Eq. (\ref{q15}) we choose the basis
$|n_0,n_1,...,n_N\rangle_d$.
From Eq. (\ref{q14}) and its hermitian conjugate we have

\bqn
\hat{a}_\mu'^\dag(t)\hat{a}_\mu'(t)&=&\sum_{\nu,\rho=0}^Nf_{\mu\rho}^\ast(t)
f_{\mu\nu}(t)\hat{a}_\rho'^\dag\hat{a}_\nu'
\nonumber\\
&=&\sum_{\nu=0}^N|f_{\mu\nu}(t)|^2
\hat{a}'^\dag_\nu \hat{a}_\nu'+
\sum_{\nu\neq\rho}f_{\mu\rho}^\ast(t)f_{\mu\nu}(t)
\hat{a}'^\dag_\nu\hat{a}_\rho'\;.
\label{q16}
\eqn
In the basis $|n_0,n_1,...,n_N\rangle_d$ the second term in the
above equation gives no contribution for Eq. (\ref{q15}). Then,
replacing Eq. (\ref{q16})  in Eq. (\ref{q15}) we obtain easily,

\be
n_\mu(t)=|f_{\mu 0}(t)|^2n_0(0)+\sum_{k=1}^N|f_{\mu k}(t)|^2n_k(0)\;,
\label{q17}
\ee
where the initial distributions for the dressed atom and field
modes are given respectively by

\bqn
n_0(0)&=&{\rm Tr}_0\left(\hat{a}_0'^\dag\hat{a}_0'\hat{\rho}_0\right)\nonumber\\
&=&\sum_{n=0}^\infty n~_d\langle n |\hat{\rho}_0|n \rangle_d
\label{q17a}
\eqn
and

\bqn n_k(0)&=&\frac{{\rm Tr}_k\left(\hat{a}_k'^\dag\hat{a}_k'
e^{-\beta\omega_k(\hat{a}_k'^\dag\hat{a}_k'+1/2)}\right)} {{\rm
Tr}_k\left(e^{-\beta\omega_k(\hat{a}_k'^\dag\hat{a}_k'+1/2)}\right)}
\nonumber\\
&=&\frac{1}{e^{\beta\omega_k}-1}\;.
\label{q17b}
\eqn

Setting $\mu=0$ in Eq. (\ref{q17}), we obtain for the time
dependent thermal expectation value of the occupation number
operator, corresponding to the atom,

\be
n_{0}(t)=|f_{0 0}(t)|^2n_0(0)+\sum_{k=1}^N|f_{0 k}(t)|^2n_k(0)\;.
\label{q18}
\ee
In early references it has been showed that $|f_{00}(t)|^2$ is the
probability of the atom to remain at time $t$ in the first excited
level, whereas $|f_{0k}|^2$ is the probability decay of the atom
from the first excited level to the ground state by emission of a
field quanta of frequency $\omega_k$
\cite{adolfo1,adolfo2,gabriel}. Then, Eq. (\ref{q18}) suggest a
clear physical interpretation in terms of these probabilities.
Also Eq. (\ref{q17}) can be interpreted in the same way.

For the frequency field modes given in the paragraph after Eq.
(\ref{a3}) and in the continuum limit $L\to\infty$ the
coefficients $f_{00}(t)$ and $f_{0k}(t)$ are calculated in
Appendix . We obtain the following values [Eqs. (\ref{abog21}) and
(\ref{abog23})]

\be
f_{00}(t)=(1-\frac{i\pi g}{2\kappa})e^{-i\kappa t-\pi gt/2}
+2ig J(t) \label{q19}
\ee
and \be f_{0k}(t)=
\sqrt{2g\Delta\omega}~\!\omega_k\left[\frac{(1-\frac{i\pi
g}{2\kappa})e^{-i\kappa t-\pi gt/2}}
{[\omega_k^2-(\kappa-\frac{i\pi g}{2})^2]} -\frac{e^{-i\omega_k
t}}{[\omega_k^2-\omega_0^2+ i\pi
g\omega_k]}\right]+2ig\sqrt{2g\Delta\omega}~\!\omega_k
I(\omega_k,t)\;, \label{q20} \ee
where $\kappa=\sqrt{\omega_0^2-\frac{\pi^2}{4} g^2}$,

\be
J(t)=\int_0^\infty dy \frac{y^2e^{-yt}}{(y^2+\omega_0^2)^2
-\pi^2 g^2y^2} \label{q21}
\ee
and

\be
I(\omega_k,t)=\int_{0}^\infty dy\frac{y^2e^{-yt}}
{[(y^2+\omega_0^2)^2-\pi^2g^2y^2](y^2+\omega_k^2)}\;. \label{q22}
\ee
Replacing Eqs. (\ref{q19}) and (\ref{q20}) in Eq. (\ref{q17}) we
obtain in the continuum limit $\Delta \omega\to 0,~N\to\infty$,

\be
n_0(t)=P_{00}(t)n_0(0) +\int_0^\infty d\omega
\frac{P_{0\omega}(t)} {\left(e^{\beta\omega}-1\right)}\;,
\label{q23}
\ee
where

\be
P_{00}(t)=\frac{\omega_0^2}{\kappa^2} e^{-\pi
gt}-2gJ(t)e^{-\pi gt/2} \left[2\sin(\kappa t)+\frac{\pi
g}{\kappa}\cos(\kappa t)\right]+4g^2J^2(t)\;, \label{q24} \ee
\bqn
P_{0\omega}(t)&=&2\frac{g}{\kappa}\omega^2\left\{\frac{\kappa^2+\omega_0^2e^{-\pi
gt}} {\kappa K(\omega)}-\frac{e^{-\pi gt/2}}{K^2(\omega)}
\Big(\left[2\kappa(\omega^2-\omega_0^2)^2+
\pi^2g^2\omega(\omega^2+\omega_0^2)\right]\cos[(\omega-\kappa)t]\right.
\nonumber\\
& & +\pi
g(\omega^2-\omega_0^2)(\omega^2+\omega_0^2-2\kappa\omega)\sin[(\omega-\kappa)t]
+2gI(\omega,t)K(\omega)\left[2\kappa(\omega^2-\omega_0^2)\sin(\kappa
t)\right.\nonumber\\
& &+\left.\pi g(\omega^2+\omega_0^2)\cos(\kappa
t)\right]\Big)+\left.4g\kappa\frac{ I(\omega,t)}{K(\omega)}
\left[(\omega^2-\omega_0^2)\sin(\omega t)+\pi g\omega\cos(\omega
t)\right] +4\kappa g^2I^2(\omega,t)\right\}\;, \label{q25} \eqn
and

\be
K(\omega)=(\omega^2-\omega_0^2)^2+\pi^2g^2\omega^2\;.
\label{q26}
\ee

Note that in the limit $t\to\infty$ Eq. (\ref{q23}) have a well
defined limit, that is, the atom reaches a final equilibrium
state. Also, in this limit the term $P_{00}(t)$, proportional to
$n_0(0)$ vanishes, that is, the final equilibrium distribution is
independent of the initial atom density operator, $\hat{\rho}_0$,
it depends exclusively on the thermal field degrees of freedom. As
showed in Refs. \cite{adolfo1,adolfo2,gabriel} $P_{00}(t)$ goes to
zero almost exponentially in a time of the order $\pi/g$. Taking
$t\to\infty$ in Eq. (\ref{q23}) we get

\be
n_0(\infty)=2g\int_0^\infty d\omega\frac{\omega^2}
{[(\omega^2-\omega_0)^2+\pi^2g^2\omega^2]
(e^{\beta\omega}-1)}\;.
\label{q27}
\ee
Now, the question is about the physical meaning of the equilibrium value
given by Eq. (\ref{q27}). To answer this question we compute
the thermal expectation value of the number operator
$\hat{a}_0'^\dag\hat{a}_0'$, in the case in which the
atom-electromagnetic field system is at thermal equilibrium at some
given temperature $\theta^{-1}$. In this case the density operator is given by

\be
\hat{\rho}_\theta=\frac{e^{-\theta\hat{H}}}{{\rm Tr}(e^{-\theta\hat{H}})}\;,
\label{q28}
\ee
where $\hat{H}$ is given by Eq. (\ref{a3}). We want to compute

\be
n_0=\frac{{\rm Tr}(\hat{a}_0'^\dag\hat{a}_0'e^{-\theta\hat{H}})}
{{\rm Tr}(e^{-\theta\hat{H}})}\;.
\label{q29}
\ee
To compute above expression we write $\hat{H}$, as

\be
\hat{H}=\sum_{r=0}^N\left(\hat{A}_r^\dag\hat{A}_r+\frac{1}{2}\right)\Omega_r\;,
\label{q30}
\ee
where $\hat{A}_r$ and $\hat{A}_r^\dag$ are the normal annihilation and creation
operators and given by

\bqn
\hat{A}_r&=&
\frac{1}{\sqrt{2\Omega_r}}\hat{P}_r-i\sqrt{\frac{\Omega_r}{2}}\hat{Q}_r
\label{q31}\\
\hat{A}_r^\dag&=&\frac{1}{\sqrt{2\Omega_r}}\hat{P}_r
+i\sqrt{\frac{\Omega_r}{2}}\hat{Q}_r\;.
\label{q32}
\eqn
Now, using Eq. (\ref{dress}) and from Eqs. (\ref{a1})-(\ref{q4}) and
(\ref{q31})-(\ref{q32}) we find that

\be
\hat{a}_\mu'=\sum_{r=0}^N t_\mu^r\hat{A}_r\;,~~~
\hat{a}_\mu'^\dag=\sum_{r=0}^N t_\mu^r\hat{A}_r^\dag\;.
\label{q33}
\ee
Using above expressions in Eq. (\ref{q29}) and computing the trace by
using the  basis $|n_0,n_1,...,n_N\rangle_c$, that are eigenvectors of
$\hat{H}$, we find easily,

\be
n_0=\sum_{r=0}^N\frac{(t_0^r)^2}{e^{\theta\Omega_r}-1}
\label{q34}
\ee
and in the continuum limit we get [see Appendix, Eq. (\ref{abog24})

\be
n_0=2g\int_0^\infty dx\frac{x^2}
{[(x^2-\omega_0)^2+\pi^2g^2x^2]
(e^{\theta x}-1)}\;.
\label{q35}
\ee
In the case in which $\theta=\beta$, Eqs. (\ref{q35}) and
(\ref{q27}) are identical. Then, we conclude from above
calculations that the atom reaches a final thermal equilibrium
distribution, it thermalizes with the thermal radiation field at
temperature $\beta^{-1}$. Note that for weak coupling
$g\ll\omega_0$, we can obtain from Eq. (\ref{q27}) or (\ref{q35}),
\be
n(\infty)\approx \frac{1}{e^{\beta\omega_0}-1}\;,
\label{q36}
\ee
a Bose-Einstein distribution, an expected textbook result.

\section{conclusions}
In this work we have showed that an atom  (approximated by the
dressed harmonic oscillator) initially in any arbitrary state and
suddenly coupled to a thermal radiation field, evolves in time to
a final thermal equilibrium state. The mean time, necessary for
this to occur, can be roughly estimated from Eqs.
(\ref{q24})-(\ref{q25}) and is of the order $\pi/g$, an
intuitively expected result. Also, we have found a physically
suggestive result for the time evolution of the thermal
expectation value of the dressed occupation number operators, Eqs.
(\ref{q17}) and Eq. (\ref{q18}). In general, this time evolution
is given in terms of the time dependent probabilities associated
with the emission and absorption of field quanta.

 \section*{Acknowledgements}
GFH is supported by FAPEMIG (Funda\c{c}\~ao de Amparo \`a Pesquisa
do Estado de Minas Gerais).

\appendix
\section{The continuum limit}

We want to compute, in the continuum limit, sums of the type

\begin{equation}
R_{\mu\nu}=\sum_{r=0}^N t_\mu^r t_\nu^r {\cal R}_{\mu\nu}(\Omega_r)\;,
\label{cc1}
\end{equation}
where ${\cal R}_{\mu\nu}(\Omega)$ is an analytic function of $\Omega$.
For this end we define a function $W(z)$,

\begin{equation}
W(z)=z^2-\omega_0^2+\sum_{k}^N\frac{\eta^2z^2}
{\omega_k^2-z^2}\;.
\label{cc5}
\end{equation}
From Eqs. (\ref{tkrg1}) and (\ref{la5b}) we can note that the
$\Omega_r$'s are the roots of $w(z)$. For complex values
of $z$ and using $\eta^2=2g\Delta \omega$, we can write
Eq. (\ref{cc5}) in the continuum limit as,

\begin{equation}
W(z)=z^2-\omega_0^2+2gz^2\int_{0}^{\infty}\frac{d\omega}
{\omega^2-z^2}\;.
\label{cc6}
\end{equation}
For complex values of $z$ the above integral is well
defined and can be evaluated easily using Cauchy theorem,
obtaining

\begin{equation}
W(z)=\left\{\begin{array}{c}
z^2+ig\pi z-\omega_0^2,~{\rm Im}(z)>0\\
z^2-ig\pi z-\omega_0^2,~{\rm Im}(z)<0\;.
\end{array}\right.
\label{cc7}
\end{equation}
We start by computing $R_{00}(t)$,

\begin{equation}
R_{00}=\sum_{r=0}^N(t_0^r)^2{\cal R}_{00}(\Omega_r)\;.
\label{cc8}
\end{equation}
From the expression for $t_0^r$, given in (\ref{tkrg1}) and Eq. (\ref{cc5}) it is
easy to show that,

\begin{equation}
(t_0^r)^2=\frac{2\Omega_r}{W'(\Omega_r)}\;,
\label{cc9}
\end{equation}
where the prime means derivative with respect to the
argument.  Since the $\Omega_r$'s are the roots of $W(z)$, we can
write Eq. (\ref{cc8}) as

\begin{equation}
R_{00}=\frac{1}{i\pi}\oint_C dz\frac{z {\cal R}_{00}(z)}{W(z)}\;,
\label{cc10}
\end{equation}
where $C$ is a counterclockwise contour in the $z$-plane
that encircles the real positive roots $\Omega_r$, that is,
a contour that encircles the real positive axis. The integral
in Eq. (\ref{cc10}) can be evaluated choosing a contour
that lies just below and above of the real positive axis.
Below the real positive axis we have $z=\alpha-i\epsilon$
and above $z=\alpha+i\epsilon$, where $\alpha$ is real positive
and $\epsilon\to 0^+$. Then, we have for Eq. (\ref{cc10}),

\begin{equation}
R_{00}=\frac{1}{i\pi}\int_{0}^{\infty}d\alpha
\left[\frac{(\alpha-i\epsilon) {\cal R}_{00}
(\alpha-i\epsilon)}{W(\alpha-i\epsilon)}-
\frac{(\alpha+i\epsilon) {\cal R}_{00}(\alpha+i\epsilon)}
{W(\alpha+i\epsilon)}\right]\;.
\label{cc11}
\end{equation}
From Eq. (\ref{cc7}) we get for $W(\alpha-i\epsilon)$ and
$W(\alpha-i\epsilon)$ respectively in the
limit $\epsilon\to 0^+$,

\begin{eqnarray}
W(\alpha+i\epsilon)&=&\alpha^2-\omega_0^2+ig\pi\alpha\;,
\nonumber\\
W(\alpha-i\epsilon)&=&\alpha^2-\omega_0^2-ig\pi\alpha\;.
\label{cc15}
\end{eqnarray}
Taking the limit $\epsilon\to 0^+$ in Eq.  (\ref{cc11}) and using
Eq. (\ref{cc15}) we get

\begin{equation}
R_{00}=2g\int_{0}^{\infty} d\alpha
\frac{\alpha^2 {\cal R}_{00}(\alpha)}{(\alpha^2-\omega_0^2)^2+
g^2\pi^2\alpha^2}\;.
\label{cc12}
\end{equation}
As a check that Eq. (\ref{cc12}) is correct we take the case ${\cal R}_{00}=1$
and using Cauchy theorem it is easy to show that above
integral is $1$, as expected from the orthonormality property of the matrix
$\{t_\mu^r\}$.

Next we compute $R_{0k}(t)$,

\begin{equation}
R_{0k}=\sum_{r=0}^N t_0^rt_k^r {\cal R}_{0k}(\Omega_r)\;.
\label{cc13}
\end{equation}
Using the expressions for $t_0^r$ and $t_k^r$, as given
by (\ref{tkrg1}), in Eq. (\ref{cc13}) we obtain

\begin{eqnarray}
R_{0k}&=&\eta\omega_k\sum_{r=0}^N\frac{(t_0^r)^2 {\cal R}_{0k}(\Omega_r)}
{(\omega_k^2-\Omega_r^2)}\nonumber\\
&=&\frac{\eta\omega_k}{i\pi}\oint_C dz\frac{z{\cal R}_{0k}(z)}
{(\omega_k^2-z^2)W(z)}\;,
\label{cc14}
\end{eqnarray}
where in the second line the pole at $z=\omega_k$ gives a zero
contribution since $W(\omega_k)$ as given by Eq. (\ref{cc5})
or (\ref{cc6}) is infinity. Evaluating Eq. (\ref{cc14})
by choosing the same contour as in the evaluation of $R_{00}(t)$
we get

\begin{equation}
R_{0k}=-\frac{\eta\omega_k}{i\pi}\int_0^{\infty}d\alpha
\left[\frac{(\alpha-i\epsilon) {\cal R}_{0k}(\alpha-i\epsilon)}
{W(\alpha-i\epsilon)[(\alpha-i\epsilon)^2-\omega_k^2]}-
\frac{(\alpha+i\epsilon) {\cal R}_{0k}(\alpha+i\epsilon)}
{W(\alpha+i\epsilon)[(\alpha+i\epsilon)^2-\omega_k^2]}\right]\;.
\label{cc16}
\end{equation}
Using Eq. (\ref{cc15}) in Eq. (\ref{cc16}) we obtain,

\begin{eqnarray}
R_{0k}&=&-\frac{\eta\omega_k}{i\pi}\int_0^{\infty}d\alpha
\left[\frac{\alpha {\cal R}_{0k}(\alpha)}{(\alpha-\frac{i\pi g}{2}-\kappa)
(\alpha-\frac{i\pi g}{2}+\kappa)(\alpha-i\epsilon-\omega_k)
(\alpha-i\epsilon+\omega_k)}\right.\nonumber\\
& &~~~~~~~~~~~\left.-\frac{\alpha {\cal R}_{0k}(\alpha)}
{(\alpha+\frac{i\pi g}{2}-\kappa)(\alpha+\frac{i\pi g}{2}+\kappa)
(\alpha+i\epsilon-\omega_k)(\alpha+i\epsilon+\omega_k)}
\right]\;,
\label{cc17}
\end{eqnarray}
where $\kappa=\sqrt{\omega_0^2-\frac{\pi^2}{4}g^2}$. To check the
validity of Eq. (\ref{cc17}) we take
${\cal R}_{0k}=1$ and using Cauchy theorem it can be proved that the integral
vanishes as expected from the orthonormality of the matrix $\{t_\mu^r\}$.

Now, it is straightforward to compute the coefficients $f_{\mu\nu}(t)$

\be
f_{\mu\nu}(t)=\sum_{r=0}^Nt_\mu^rt_\nu^re^{-it\Omega_r}
\label{abog19}
\ee
in the continuum limit.
Taking $\mu=\nu=0$ in Eq. (\ref{abog19}) and using Eq. (\ref{cc12}) we get

\be
f_{00}(t)=2g\int_{0}^{\infty} dx
\frac{x^2 e^{-itx}}{(x^2-\omega_0^2)^2+
g^2\pi^2x^2}\;,
\label{abog20}
\ee
from which we find

\begin{equation}
f_{00}(t)=(1-\frac{i\pi g}{2\kappa})e^{-i\kappa t-\pi gt/2}
+2ig\int_0^\infty dy \frac{y^2e^{-yt}}{(y^2+\omega_0^2)^2
-\pi^2 g^2y^2}\;,~~~~~(\kappa^2>0)\;.
\label{abog21}
\end{equation}
Taking $\mu=0,~\nu=k$ in Eq. (\ref{abog19}) and using Eq. (\ref{cc17})
we get

\begin{eqnarray}
f_{0k}&=&-\frac{\eta\omega_k}{i\pi}\int_0^{\infty}dx
\left[\frac{x e^{-itx}}{(x-\frac{i\pi g}{2}-\kappa)
(x-\frac{i\pi g}{2}+\kappa)(x-i\epsilon-\omega_k)
(x-i\epsilon+\omega_k)}\right.\nonumber\\
& &~~~~~~~~~~~\left.-\frac{x e^{-itx}}
{(x+\frac{i\pi g}{2}-\kappa)(x+\frac{i\pi g}{2}+\kappa)
(x+i\epsilon-\omega_k)(x+i\epsilon+\omega_k)}
\right]\;.
\label{abog22}
\end{eqnarray}
We can integrate Eq. (\ref{abog22}) in the complex plane by using
cauchy theorem. We choice as the closed contour of integration,
the path that goes in the real axis from $0$ to $\infty$, then go
to the negative imaginary axis along the part of the circle with
radius $R\to\infty$ and argument $-\pi/2<\theta<0$ and closes the
contour along the imaginary axis from $-i\infty$ to the origin.
Note that inside the contour of integration only the second term
in the bracket of Eq. (\ref{cc17}) has two poles at
$-ig\pi/2+\kappa$ and $-i\epsilon+\omega_k$. Then, we get for Eq.
(\ref{abog22})

\begin{eqnarray}
f_{0k}(t)&=&-\frac{\eta\omega_k}{i\pi}\left\{(-2i\pi)\left[
\frac{ (\frac{i\pi g}{2\kappa}-1)e^{-i\kappa t-\pi gt/2}}
{2[(\kappa-\frac{i\pi g}{2})^2-\omega_k^2]}-\frac{e^{-i\omega_kt}}
{2[\omega_k^2+i\pi g\omega_k-\omega_0^2]}\right]\right.\nonumber\\
& &-\left.\int_{-\infty}^0 dy\left[\frac{ye^{yt}}{(-y^2+\pi gy
-\omega_0^2)(y^2+\omega_k^2)}-\frac{ye^{yt}}{(-y^2-\pi gy
-\omega_0^2)(y^2+\omega_k^2)}\right]\right\}\nonumber\\
&=&\eta\omega_k\left[\frac{(1-\frac{i\pi g}{2\kappa})e^{-i\kappa t
-\pi gt/2}}{[\omega_k^2-(\kappa-\frac{i\pi g}{2})^2]}
-\frac{e^{-i\omega_k t}}{[\omega_k^2-\omega_0^2+
i\pi g\omega_k]}\right]\nonumber\\
& &~~~~~~~~~~~~~~~+2ig\eta\omega_k\int_{0}^\infty dy\frac{y^2e^{-yt}}
{[(y^2+\omega_0^2)^2-\pi^2g^2y^2](y^2+\omega_k^2)}\;,~~~~~(\kappa^2>0)\;.
\label{abog23}
\end{eqnarray}

To compute

\be
n_0=\sum_{r=0}^N\frac{(t_0^r)^2}{e^{\theta\Omega_r}-1}
\label{abog24}
\ee
in the continuum limit we use Eq. (\ref{cc12}), obtaining

\be
n_0=2g\int_0^\infty dx\frac{x^2}
{[(x^2-\omega_0)^2+\pi^2g^2x^2]
(e^{\theta x}-1)}\;.
\label{abog25}
\ee




\begin{thebibliography}{99}

\bibitem{reviews} J. S. Langer in {\it Solids Far from Equilibrium}, Ed. C.
Godr\`eche (Cambridge University Press, Cambridge, 1992); D. J. Evans and G. P.
Morriss, {\it Statistical Mechanics of Non-Equilibrium Liquids}
(Academic Press, London, 1990).

\bibitem{DCC} K. Rajagopal and F. Wilczek, Nucl. Phys. {\bf B 399}, 395 (1995);
ibid. {\bf B 404} (1993) 577.

\bibitem{lindereh}L. Kofman, A. Linde and A. A. Starobinsky,
Phys. Rev. {\bf D 56} (1997) 3258.

\bibitem{GR}M. Gleiser and R. O. Ramos, Phys. Rev. {\bf D 50} (1994) 2441.

\bibitem{BGR}A. Berera, M. Gleiser and R. O. Ramos, Phys. Rev. {\bf D 58}
(1998) 123508.

\bibitem{BR}A. Berera and R. O. Ramos, Phys.
Rev. {\bf D 63} (2001) 103509.

\bibitem{BEC}D. G. Barci, E. S. Fraga and R. O. Ramos,
Phys. Rev. Lett. {\bf 85} (2000) 479;
Laser Phys.  {\bf 12} (2002) 43;
D. G. Barci, E. S. Fraga, M. Gleiser and R. O. Ramos,
Physica {\bf A 317} (2003) 535.

\bibitem{RF}R. O. Ramos and F. A. R. Navarro, Phys. Rev. {\bf D 62}
(2000) 085016.

\bibitem{salle} M. Salle, J. Smit and J. C. Vink, Phys. Rev. D{\bf 64},
025016 (2001).

\bibitem{parisi} G. Parisi, Europhys. Lett. {\bf 40}, 357 (1997).

\bibitem{aarts1} G. Aarts, G. F. Bonini and Ch. Wetterich, Nucl. Phys.
B{\bf 587}, 403 (2000).

\bibitem{aarts2} G. Aarts, G. F. Bonini and Ch. Wetterich,
Phys. Rev. D{\bf 63}, 025012 (2001).

\bibitem{mosko} M. Mosko and V. Cambel, Phys. Rev. B{\bf 50}, 8864 (1994).

\bibitem{srednicki} M. Srednicki, Phys. Rev. E{\bf 50}, 888 (1994).

\bibitem{berry} M. V. Berry, J. Phys. A{\bf 10}, 2083 (1977).

\bibitem{gemmer} J. Gemmer, A. Otte and G. Mahler, Phys. Rev. Lett.
{\bf 86}, 1927 (2001).

\bibitem{tasaki} H. Tasaki, Phys. Rev. Lett. {\bf 80}, 1373 (1998).

\bibitem{scarani} V. Scarani, M. Ziman, P. Stelmachovic, N. Gisin and V. Buzek,
Phys. Rev. Lett. {\bf 88}, 097905 (2002).


\bibitem{adolfo1} N. P. Andion, A.P.C. Malbouisson and A. Mattos Neto, J.Phys.
A{\bf 34}, 3735 (2001).

\bibitem{adolfo2} G. Flores-Hidalgo, A.P.C. Malbouisson and Y.W. Milla,  Phys. Rev. A,
{\bf 65}, 063414 (2002), arXiv:physics/0111042.

\bibitem{gabriel} G. Flores-Hidalgo and A.P.C. Malbouisson, Phys. Rev. A{\bf66}, 042118
(2002), arXiv:quant-ph/0205042.

\bibitem{hulet} R. G. Hulet, E. S. Hilfer, D. Kleppner,
Phys. Rev. Lett. {\bf 55}, 2137 (1985).

\bibitem{haroche2} W. Jhe, A. Anderson, E. A. Hinds, D. Meschede, L. Moi and
S. Haroche, Phys. Rev. Lett. {\bf 58}, 666 (1987).

\bibitem{nonlinear} G. Flores-Hidalgo and A. P. C. Malbouisson,
Phys. Lett. A{\bf 311}, 82 (2003), arXiv:physics/0211123.

\bibitem{yony} G. Flores-Hidalgo and Y. W. Milla,
 J. Phys. A: Math. Gen. {\bf 38}, 7527 (2005), arXiv:physics/0410238.

\bibitem{casana} R. Casana, G. Flores-Hidalgo and B. M. Pimentel,
Phys. Lett. A{\bf 337}, 1 (2005), arXiv:physics/0410063.

\bibitem{eletromag} G. Flores-Hidalgo and A. P. C.
Malbouisson, Phys. Lett. A{\bf 337}, 37 (2005),
arXiv:physics/0312003.

\bibitem{casana2}R. Casana, G. Flores-Hidalgo and B. M. Pimentel,
Physica A{\bf 374}, 600 (2007), arXiv: physics/0506223.

\bibitem{thirring}  W. Thirring and F. Schwabl, Ergeb. Exakt. Naturw. {\bf 36},
219 (1964).

\bibitem{weiss} U. Weiss, {\it Quantum dissipative systems}, (World Scientific
Publishing Co. Singapore 1993).


\bibitem{feynman} R. P. Feynman and F. L Vernon, Ann. Phys. (NY) {\bf 24}, 118
(1963); {\it ibid} {\bf 281}, 547 (2000).

\bibitem{ulersma} P. Ullersma, Physica {\bf 32}, 27 (1966).

\bibitem{caldeira} A. O. Caldeira and A. J. Leggett, Physica A{\bf 121},
587 (1983).

\bibitem{zurek1} W. G. Unruh and W. H. Zurek, Phys. Rev. D{\bf40}, 1071 (1989).

\bibitem{zurek2} W. H. Zurek, Phys. Today, {\bf 44}, 36 (1991).

\bibitem{paz} B. L. Hu, J. P. Paz and Y. Zhang, Phys, Rev. D, {\bf 45},
2843 (1992).

\bibitem{gabrielrudnei} G. Flores-Hidalgo and R. O. Ramos,
Physica A{\bf 326}, 159 (2003), arXiv:hep-th/0206022.


\end{thebibliography}
\end{document}